\documentstyle[aps,epsfig]{revtex}
\begin{document}

\title{Folding model analysis of proton radioactivity of spherical proton emitters}

\author{D.N. Basu$^1$\thanks{E-mail:dnb@veccal.ernet.in}, P. Roy Chowdhury$^2$, C. Samanta$^{2,3}$}
\address{$^1$ Variable  Energy  Cyclotron  Centre,  1/AF Bidhan Nagar, Kolkata 700 064, India}
\address{ $^2$ Saha Institute of Nuclear Physics, 1/AF Bidhan Nagar, Kolkata 700 064, India }
\address{ $^3$ Physics Department, Virginia Commonwealth University, Richmond, VA 23284-2000, U.S.A. }

\date{\today }
\maketitle
\begin{abstract}

      Half lives of the decays of spherical nuclei away from proton drip line by proton emissions are estimated theoretically. The quantum mechanical tunneling probability is calculated within the WKB approximation. Microscopic proton-nucleus interaction potentials are obtained by single folding the densities of the daughter nuclei with M3Y effective interaction supplemented by a zero-range pseudo-potential for exchange along with the density dependence. Parameters of the density dependence are obtained from the nuclear matter calculations. Spherical charge distributions are used for Coulomb interaction potentials. These calculations provide reasonable estimates for the observed proton radioactivity lifetimes of proton rich nuclei for proton emissions from 26 ground and isomeric states of spherical proton emitters. 



\end{abstract}



\section{INTRODUCTION}
\label{sectoin1}

      The proton separation energies of nuclei lying in the domain beyond the proton drip line are negative. Consequently these proton rich nuclei have positive $Q$ values for proton emissions with a natural tendency to shed off excess protons and are spontaneous proton emitters. The phenomenon of proton emission from nuclear ground states limits the possibilities of the creation of more exotic proton rich nuclei which are usually produced by fusion-evaporation nuclear reactions. Apart from providing the limit to the proton dripline, the one proton radioactivity may be used as a tool to obtain spectroscopic information because the decaying proton is the unpaired proton not filling its orbit. These decay rates are sensitive to the $Q$ values and the orbital angular momenta which in turn help to determine the orbital angular momenta of the emitted protons.  

      Since the observation of proton radioactivity is comparatively recent, only few theoretical attempts have been made to study this exotic process \cite{r1,r2,r3,r4}. In the energy domain of radioactivity, proton can be considered as a point charge having highest probability of being present in the parent nucleus. It has the lowest Coulomb potential among all charged particles and mass being smallest it suffers the highest centrifugal barrier, enabling this process suitable to be dealt within WKB barrier penetration model. In the existing theoretical models \cite{r1,r2} for proton radioactivity, Saxon-Woods type potential has been used for the nuclear interaction. In another recent work \cite{r4}, a unified fission model with proximity potential for nuclear force has been used. In the present work, quantum mechanical tunneling probability is calculated within the WKB approximation using microscopic proton-nucleus interaction potentials. These potentials have been obtained by single folding the densities of daughter nuclei with a realistic effective interaction supplemented by a zero-range pseudo-potential for exchange along with density dependence. Calculations using such potentials provide excellent estimates for lifetimes of the exotic decay process of proton radioactivity. 

      A well-defined effective nucleon-nucleon (NN) interaction in the nuclear medium is important not only for different structure models but also for the microscopic calculation of the nucleon-nucleus and nucleus-nucleus potentials used in the analysis of the nucleon and heavy-ion scattering. Effective NN interaction can be best constructed from a sophisticated G-matrix calculation. This interaction has been derived by fitting its matrix elements in an oscillator basis to those elements of the G-matrix obtained with the Reid-Elliott soft-core NN interaction \cite{r5}. The ranges of the M3Y forces were chosen to ensure a long-range tail of the one-pion exchange potential as well as a short range repulsive part simulating the exchange of heavier mesons. Such an effective NN interaction has been shown to provide a more realistic shape of the scattering potentials of the nucleon or heavy ion optical potentials obtained by folding in the density distribution functions of two interacting nuclei with the effective NN interaction \cite{r6}.

      The density dependent M3Y (DDM3Y) effective NN interaction has been used to determine the incompressibility of infinite nuclear matter \cite{r7}. The equilibrium density of the nuclear matter has been determined by minimising the energy per nucleon. The density dependence parameters have been extracted by reproducing the saturation energy per nucleon and the saturation density of spin and isospin symmetric cold infinite nuclear matter. Result of such calculations also provide a reasonable value of nuclear incompressibility. In nuclear matter calculations, the calculation of potential energy per nucleon involves folding of interaction of one nucleon with the rest of the nuclear matter. It is therefore used in single folding model description for nuclear matter calculations and thus density dependence parameters obtained from nuclear matter calculations may be used as it is in describing nucleon-nucleus interaction potentials where single folding model comes into play. Such nucleon-nucleus interaction potentials have been used successfully to the analysis of elastic and inelastic scattering of protons \cite{r8}. 

      In the present work we provide estimates for the proton radioactivity lifetimes of the spherical proton emitters from the ground and the isomeric states using the same nucleon-nucleus interaction potentials obtained microscopically by single folding the daughter nuclei density distributions with a realistic DDM3Y effective interaction whose density dependence parameters have been extracted from the nuclear matter calculations.
 
\section{FORMALISM}
\label{section2}

      The microscopic nuclear potentials $V_N(R)$ have been obtained by single folding the density of the daughter nucleus with the finite range realistic DDM3Y effective interacion as

\begin{equation}
 V_N(R) = \int \rho (\vec{r}) v[|\vec{r} - \vec{R}|] d^3r 
\label{seqn1}
\end{equation}
\noindent
where $\vec{R}$ and $\vec{r}$ are, respectively, the co-ordinates of the emitted proton and a nucleon belonging to the residual daughter nucleus with respect to its centre. The density distribution function $\rho$ used for the daughter nucleus, has been chosen to be of the spherically symmetric form given by

\begin{equation}
 \rho(r) = \rho_0 / [ 1 + exp( (r-c) / a ) ]
\label{seqn2}
\end{equation}                                                                                                                                           \noindent     
where                        
 
\begin{equation}
 c = r_\rho ( 1 - \pi^2 a^2 / 3 r_\rho^2 ), ~~    r_\rho = 1.13 A_d^{1/3}  ~~   and ~~    a = 0.54 ~ fm
\label{seqn3}
\end{equation}
\noindent
and the value of $\rho_0$ is fixed by equating the volume integral of the density distribution function to the mass number $A_d$ of the residual daughter nucleus. The distance s between a nucleon belonging to the residual daughter nucleus and the emitted proton is given by

\begin{equation}
 s = |\vec{r} - \vec{R}|
\label{seqn4}
\end{equation}   
\noindent
while the interaction potential between any such two nucleons $v(s)$ appearing in eqn.(1) is given by the DDM3Y effective interaction. The total interaction energy $E(R)$ between the proton and the residual daughter nucleus is equal to the sum of the nuclear interaction energy, the Coulomb interaction energy and the centrifugal barrier. Thus

\begin{equation}
 E(R) = V_N(R) + V_C(R) + \hbar^2 l(l+1) / (2\mu R^2)
\label{seqn5}
\end{equation}   
\noindent
where $\mu = M_p M_d/M_A$  is the reduced mass, $M_p$, $M_d$ and $M_A$ are the masses of the proton, the daughter nucleus and the parent nucleus respectively, all measured in the units of $MeV/c^2$. Assuming spherical charge distribution (SCD) for the residual daughter nucleus, the proton-nucleus Coulomb interaction potential $V_C(R)$ is given by

\begin{eqnarray}
 V_C(R) =&& Z_d e^2/ R~~~~~~~~~~~~~~~~~~~~~~~~~~~~~~~~~~for~~~~R \geq R_c \nonumber\\
            =&&( Z_d e^2/ 2R_c).[ 3 - (R/R_c)^2]~~~~~~~~~~for~~~~R\leq R_c 
\label{seqn6}
\end{eqnarray}   
\noindent
where $Z_d$ is the atomic number of the daughter nucleus. The touching radial separation $R_c$ between the proton and the daughter nucleus is given by $R_c = c_p+c_d$ where $c_p$ and $c_d$ have been obtained using eqn.(3). The energetics allow spontaneous emission of protons only if the released energy

\begin{equation}
 Q = [ M_A - ( M_p + M_d ) ] c^2
\label{seqn7}
\end{equation}
\noindent
is a positive quantity.

      In the present work, the half life of the parent nucleus decaying via proton emission is calculated using the WKB barrier penetration probability. The assault frequency $\nu$ is obtained from the zero point vibration energy $E_v = (1/2)\hbar\omega = (1/2)h\nu$. The decay half life $T$ of the parent nucleus $(A, Z)$  into a proton and a daughter $(A_d, Z_d)$  is given by

\begin{equation}
 T = [(h \ln2) / (2 E_v)] [1 + \exp(K)]
\label{seqn8}
\end{equation}
\noindent
where the action integral $K$ within the WKB approximation is given by

\begin{equation}
 K = (2/\hbar) \int_{R_a}^{R_b} {[2\mu (E(R) - E_v - Q)]}^{1/2} dR
\label{seqn9}
\end{equation}
\noindent
where $R_a$ and $R_b$ are the two turning points of the WKB action integral determined from the equations

\begin{equation}
 E(R_a)  = Q + E_v =  E(R_b)
\label{seqn10}
\end{equation} 
\noindent
From a fit to the experimental data on cluster emitters a law given by eqn.(5) of reference \cite{r9}, which relates $E_v$ with $Q$, was found. For the present calculations same law extended to protons is used for the zero point vibration energies. The shell effects of proton radioactivity is implicitly contained in the zero point vibration energy due to its proportionality with the $Q$ value.

\section{CALCULATIONS}
\label{section3}

      The M3Y interaction is based upon a realistic G-matrix. Since the G-matrix was constructed in an oscillator representation, it is effectively an average over a range of nuclear densities and therefore the M3Y has no explicit density dependence. For the same reason there is also an average over energy and the M3Y has no explicit energy dependence either. The only energy dependent effects that arises from its use is a rather weak one contained in an approximate treatment of single-nucleon knock-on exchange. The success of the extensive analysis \cite{r6} indicates that these two averages are adequate for the real part of the optical potential for heavy ions at energies per nucleon of $< 20MeV$. However, it is important to consider the density and energy dependence explicitly for the analysis of $\alpha$-particle scattering at higher energies ($>100 MeV$) where the effects of a nuclear rainbow are seen and hence the scattering becomes sensitive to the potential at small radii. Such cases were studied introducing suitable and semirealistic explicit density dependence \cite{r10,r11} into the M3Y interaction which was then called the DDM3Y and was very successful for interpreting consistently the high energy elastic $\alpha$ and heavy-ion scattering data. Present calculations have been performed using $v(s)$, inside the integral of eqn.(1) for the single folding procedure, as the DDM3Y effective \cite{r8} interaction given by

\begin{equation}
  v(s,\rho,E) = t^{\rm M3Y}(s,E)g(\rho,E)
\label{seqn11}
\end{equation}   
\noindent
where $t^{\rm M3Y}$ is the same M3Y interaction supplemented by a zero range pseudo-potential is 

\begin{equation}
 t^{\rm M3Y} = 7999 \frac{e^{ - 4s}}{4s} - 2134\frac{e^{- 2.5s}}{2.5s} + J_{00}(E) \delta(s)
\label{seqn12}
\end{equation}   
\noindent
where the zero-range pseudo-potential representing the single-nucleon exchange term is given by

\begin{equation}
 J_{00}(E) = -276 (1 - 0.005E / A_p ) (MeV.fm^3)
\label{seqn13}
\end{equation}   
\noindent
where $E$ and $A_p$ are the laboratory energy and projectile mass number respectively. In the present case of proton radioactivity it can be shown that $E/A_p=Q.m/\mu$ where m and $\mu$ are the nucleonic mass  and reduced mass of the $p+A_d$ system, respectively, in units of $MeV/c^2$. The density dependent part has been taken to be \cite{r11}

\begin{equation}
 g(\rho, E) = C (1 - \beta(E)\rho^{2/3}) 
\label{seqn14}
\end{equation}   
\noindent
which takes care of the higher order exchange effects and the Pauli blocking effects. Constants of this interaction $C$ and $\beta$ when used in single folding model description, can be determined from the nuclear matter calculations \cite{r7} as 2.07 and 1.624  fm$^2$ respectively. 

      The two turning points of the action integral given by eqn.(9) have been obtained by solving eqns.(10) using the microscopic single folding potential given by eqn.(1) along with the Coulomb potential given by eqn.(6) and the centrifugal barrier described in eqn.(5). Then the WKB action integral between these two turning points has been evaluated numerically using eqn.(1), eqn.(5), eqn.(6), eqn.(7) and eqn.(5) of reference \cite{r9}. Finally the half lives have been obtained using eqn.(8). 

\section{RESULTS AND DISCUSSIONS}
\label{section4}

      In this work, the same set of experimental data of reference \cite{r4} for the proton decay half lives have been chosen for comparison with the present theoretical calculations. Experimentally measured values of the released energy $Q$ (given by eqn.(7)), which is one of the crucial quantity for quantitative predictions of the decay half lives, have been used for the calculations. The proton emitters and the experimental values for their logarithmic half lives have been presented in Table-I. The corresponding results of the present calculations with microscopic potentials are also presented along with the results of the modified preformed cluster model (PCM) called the unified fission model (UFM) calculations \cite{r4}. The three turning points $R_1$, $R_2=R_a$ and $R_3=R_b$ obtained by solving eqn.(10) have been listed in the Table-I. 

      Experimentally measured and theoretically calculated half-lives of spherical proton emitters have been provided in Table-I. Positions of the turning points are very sensitive to the Coulomb barrier. Comparing the results for ground and isomeric states of same proton emitters it can be observed that the positions of the turning points are quite sensitive to the centrifugal barriers. Results of the present calculations with DDM3Y have been found to predict the general trend of the experimental data very well. The quantitative agreement with experimental data is good. The discrepancy between the results of present calculation and the experimental values for some cases may be due to the uncertainty in the measurements of the $Q$ values to which the results are quite sensitive due to its proportionality with the zero point vibration energies. The degree of reliability of the present estimates for the proton decay lifetimes are equivalent to the very recent UFM estimates. Changing the value of density dependence parameter $\beta$ to 1.668  fm$^2$ \cite{r12}, obtained from nuclear matter calculations using saturation energy per nucleon obtained from fitting the masses of Audi-Wapstra-Thibault \cite{r13} mass table, causes insignificant changes in the second decimal places of logarithmic half lives in some cases.

\begin{table}
\caption{Comparison between experimentally measured and theoretically calculated half-lives of spherical proton emitters. The asterisk symbol (*) denotes the isomeric state. The experimental $Q$ values, half lives and $l$ values are taken from reference [4]. The results of the present calculations have been compared with the experimental values and with the results of UFM estimates [4]. Experimental errors in $Q$ [14] values and corresponding errors in  calculated half-lives are given within parentheses. }

\begin{tabular}{ccccccccc}
Parent &Angular &Released &1st turning &2nd turning &3rd turning &Expt. &Present calc.&UFM       \\ 
nuclei &momentum& Energy & point(fm)& point(fm)&point(fm)&  & &\\ \hline
 & $l(\hbar)$ & $Q(MeV)$ &$R_1$&$R_2=R_a$&$R_3=R_b$ &$log_{10}T(s)$ &$log_{10}T(s)$& $log_{10}T(s)$ \\ \hline
&&&&&&&&\\
$^{105}Sb$&2&0.491(15)&1.55&6.58&134.30&2.049&1.97(46)&2.085\\ 
$^{145}Tm$&5&1.753(10)&3.49&6.40&56.27&-5.409&-5.14(6)&-5.170\\ 
$^{147}Tm$&5&1.071(3)&3.51&6.40&88.65&0.591&0.98(4)&1.095\\ 
$^{147}Tm^*$&2&1.139(5)&1.58&7.15&78.97&-3.444&-3.39(5)&-3.199\\ 
$^{150}Lu$&5&1.283(4)&3.50&6.44&78.23&-1.180&-0.58(4)&-0.859\\ 
$^{150}Lu^*$&2&1.317(15)&1.59&7.20&71.79&-4.523&-4.38(15)&-4.556\\ 
$^{151}Lu$&5&1.255(3)&3.51&6.49&78.41&-0.896&-0.67(3)&-0.573\\ 
$^{151}Lu^*$&2&1.332(10)&1.59&7.22&69.63&-4.796&-4.88(9)&-4.715\\ 
$^{155}Ta$&5&1.791(10)&3.51&6.55&57.83&-4.921&-4.65(6)&-4.637\\ 
$^{156}Ta$&2&1.028(5)&1.61&7.23&94.18&-0.620&-0.38(7)&-0.461\\ 
$^{156}Ta^*$&5&1.130(8)&3.52&6.53&90.30&0.949&1.66(10)&1.446\\ 
$^{157}Ta$&0&0.947(7)&0.00&7.42&98.95&-0.523&-0.43(11)&-0.126\\ 
$^{160}Re$&2&1.284(6)&1.62&7.30&77.67&-3.046&-3.00(6)&-3.109\\ 
$^{161}Re$&0&1.214(6)&0.00&7.48&79.33&-3.432&-3.46(7)&-3.231\\ 
$^{161}Re^*$&5&1.338(7)&3.52&6.63&77.47&-0.488&-0.60(7)&-0.458\\ 
$^{164}Ir$&5&1.844(9)&3.54&6.68&59.97&-3.959&-3.92(5)&-4.193\\ 
$^{165}Ir^*$&5&1.733(7)&3.52&6.69&62.35&-3.469&-3.51(5)&-3.428\\ 
$^{166}Ir$&2&1.168(8)&1.61&7.35&87.51&-0.824&-1.11(10)&-1.160\\ 
$^{166}Ir^*$&5&1.340(8)&3.56&6.70&80.67&-0.076&0.21(8)&0.021\\ 
$^{167}Ir$&0&1.086(6)&0.00&7.54&91.08&-0.959&-1.27(8)&-0.943\\ 
$^{167}Ir^*$&5&1.261(7)&3.53&6.72&83.82&0.875&0.69(8)&0.890\\ 
$^{171}Au$&0&1.469(17)&0.00&7.60&69.09&-4.770&-5.02(15)&-4.794\\ 
$^{171}Au^*$&5&1.718(6)&3.52&6.77&64.25&-2.654&-3.03(4)&-2.917\\ 
$^{177}Tl$&0&1.180(20)&0.00&7.62&88.25&-1.174&-1.36(25)&-0.993\\ 
$^{177}Tl^*$&5&1.986(10)&3.53&6.89&57.43&-3.347&-4.49(6)&-4.379\\ 
$^{185}Bi$&0&1.624(16)&0.00&7.77&65.71&-4.229&-5.44(13)&-5.184\\ 
 
\end{tabular} 

\end{table}

      For an interesting comparison, the entire calculations have been redone with the recent global optical model potential (GOMP) for protons \cite{r15}. The real central part of the GOMP for protons is given by

\begin{equation}
 V_{GOMP}(R) = -V_p(E) f(R) 
\label{seqn15}
\end{equation}
\noindent 
where the form factor $f(R)$ is given by 

\begin{equation}
 f(R) = 1/(1+exp[(R-R_V)/a_V]),~~ R_V=1.3039A_d^{1/3}-0.4054,~~a_V=0.6778-1.487 \times 10^{-4}A_d, 
\label{seqn16}
\end{equation}
\noindent
and the depth of the potential $V_p(E) $ is given by 

\begin{equation}
 V_p(E) = v^p_1[1-v^p_2(E-E^p_f)+v^p_3(E-E^p_f)^2-v^p_4(E-E^p_f)^3]+\Delta V_C(E)
\label{seqn17}
\end{equation}
\noindent   
where the Coulomb correction term $\Delta V_C(E)$ is given by 

\begin{equation}
 \Delta V_C(E) = \bar V_C v^p_1 [v^p_2-2v^p_3(E-E^p_f)+3v^p_4(E-E^p_f)^2]
\label{seqn18}
\end{equation}
\noindent     
with $v^p_1=59.3+21.0(A_d-2Z_d)/A_d-0.024A_d$, $v^p_2=0.007067+4.23\times10^{-6}A_d$, $v^p_3=1.729\times10^{-5}+1.136\times10^{-8}A_d$, $v^p_4=7\times10^{-9}$, $E^p_f=-8.4075+0.01378A_d$, $ \bar V_C=1.73Z_d/[r_cA_d^{1/3}]$, $r_c=1.198+0.697A_d^{-2/3}+12.994A_d^{-5/3}$. The lab energy $E=Q$ for the proton decay process. This GOMP $V_{GOMP}(R)$ in place of $V_N(R)$ of eqn.(5) along with the centrifugal and Coulomb potentials, with $R_c$ of eqn.(6) taken equal to $[r_cA_d^{1/3}]$ for the Coulomb potential, have been used to evaluate the action integral. Results of these calculations have been presented in Table-II. 

      The isovector or the symmetry component of the DDM3Y folded potential $V^{Lane}_N(R)$ \cite{r16} has been added to the isoscalar part of the folded potential whose results have already been presented in Table-I. The nuclear potential $V_N(R)$ of eqn.(5), therefore, has been replaced by $V_N(R)+V^{Lane}_N(R)$ \cite{r17} where

\begin{equation}
 V^{Lane}_N(R) = \int \int [\rho_{1n}(\vec{r_1})-\rho_{1p}(\vec{r_1})] [\rho_{2n}(\vec{r_2})-\rho_{2p}(\vec{r_2})] v_1[|\vec{r_2} - \vec{r_1} + \vec{R}|] d^3r_1 d^3r_2 
\label{seqn19}
\end{equation}
\noindent
where the subscripts 1 and 2 denote the daughter and the emitted nuclei respectively while the subscripts n and p denote neutron and proton densities respectively. With simple assumption that $\rho_{1p}=[\frac{Z_d}{A_d}]\rho$ and $\rho_{1n}=[\frac{(A_d-Z_d)}{A_d}]\rho$, and for the emitted particle being proton $\rho_{2n}(\vec{r_2})- \rho_{2p} (\vec{r_2})=-\rho_2(\vec{r_2})=-\delta(\vec{r_2})$, the Lane potential becomes  
$ V^{Lane}_N(R) = -[\frac{(A_d-2Z_d)}{A_d}] \int \rho (\vec{r}) v_1 [|\vec{r} - \vec{R}|] d^3r $ where $v_1(s)=t^{\rm M3Y}_1(s,E)g(\rho,E)$ and for the isovector part $t^{\rm M3Y}_1$ \cite{r6} is given by

\begin{equation}
 t^{\rm M3Y}_1 = -[4886 \frac{e^{ - 4s}}{4s} - 1176\frac{e^{- 2.5s}}{2.5s}] + 228 (1 - 0.005 Q.m/\mu ) \delta(s).
\label{seqn16}
\end{equation}   
\noindent
The inclusion of this Lane potential causes insignificant changes in the lifetimes as can be seen from Table-II. Although the lifetimes obtained using GOMP are rather close to that using isoscalar folded potentials with isovector Lane potentials (FMPL) but the GOMP and FMPL are quite different at 1st and 2nd turning points while at 3rd turning points only Coulomb potentials and centrifugal barriers are effective and nuclear potentials are negligibly small.

\begin{table}
\caption{Comparison between theoretically calculated half-lives of spherical proton emitters using the GOMP [15] and FMPL respectively. The asterisk symbol (*) denotes the isomeric state. Experimental $Q$ values and $l$ values used are taken from reference [4]. Errors in calculated half-lives arising out of experimental errors in $Q$ [14] values are given within parentheses. The overall normalization constant C=2.07 is not included in FMPL listed below at the turnings points and they should be multiplied by C to obtain their values used in the calculations or comparing them with the GOMP. }

\begin{tabular}{ccccccccccccc}
Parent &1st  &Nuclear& 2nd  &Nuclear& 3rd  & GOMP &1st &Nuclear&2nd &Nuclear&3rd & FMPL          \\ 
 &turning&GOMP&turning&GOMP&turning&&turning&FMPL&turning&FMPL&turning&\\
nuclei&point(fm) &at $R_1$& point(fm) &at $R_2$& point(fm) &  &point(fm) &at $R_1$& point(fm) &at $R_2$& point(fm) &  \\ \hline
&$R_1$&MeV&$R_2=R_a$&MeV&$R_3=R_b$&$log_{10}T(s)$&$R_1$&MeV&$R_2=R_a$&MeV&$R_3=R_b$&$log_{10}T(s)$ \\ \hline

$^{105}Sb$&1.72&-60.5&6.58&-13.2&134.30&1.97(45)&1.52&-36.1&6.61&-6.4&134.30&1.95(46)\\ 
$^{145}Tm$&3.89&-59.9&6.56&-27.4&56.27&-5.23(6)&3.43&-35.7&6.47&-13.5&56.27&-5.18(6)\\ 
$^{147}Tm$&3.90&-60.5&6.60&-27.7&88.65&0.86(3)&3.41&-36.1&6.46&-14.0&88.65&0.94(4)\\ 
$^{147}Tm^*$&1.77&-61.6&7.25&-14.2&78.97&-3.44(5)&1.54&-36.5&7.19&-7.1&78.97&-3.41(5)\\ 
$^{150}Lu$&3.93&-60.4&6.64&-27.7&78.23&-0.70(4)&3.44&-35.9&6.51&-13.9&78.23&-0.63(4)\\ 
$^{150}Lu^*$&1.78&-61.5&7.27&-14.7&71.79&-4.43(14)&1.55&-36.3&7.24&-7.0&71.79&-4.40(15)\\ 
$^{151}Lu$&3.91&-60.6&6.66&-27.7&78.41&-0.78(3)&3.44&-36.0&6.52&-13.9&78.41&-0.70(3)\\ 
$^{151}Lu^*$&1.79&-61.6&7.29&-14.7&69.63&-4.93(9)&1.56&-36.5&7.25&-7.0&69.63&-4.90(9)\\ 
$^{155}Ta$&3.91&-60.5&6.75&-26.9&57.83&-4.77(6)&3.44&-36.0&6.62&-13.5&57.83&-4.70(6)\\ 
$^{156}Ta$&1.81&-61.9&7.33&-15.2&94.18&-0.44(7)&1.57&-36.7&7.27&-7.5&94.18&-0.41(7)\\ 
$^{156}Ta^*$&3.92&-60.9&6.73&-27.9&90.30&1.53(10)&3.45&-36.2&6.60&-14.0&90.30&1.61(10)\\ 
$^{157}Ta$&0.00&-62.1&7.52&-12.5&98.95&-0.49(11)&0.00&-35.8&7.48&-6.0&98.95&-0.46(11)\\ 
$^{160}Re$&1.79&-61.8&7.40&-15.1&77.67&-3.06(6)&1.59&-36.8&7.33&-7.4&77.67&-3.02(6)\\ 
$^{161}Re$&0.00&-62.0&7.58&-12.4&79.33&-3.52(7)&0.00&-35.8&7.51&-6.1&79.33&-3.48(7)\\ 
$^{161}Re^*$&3.93&-61.0&6.84&-27.1&77.47&-0.73(7)&3.45&-36.3&6.70&-13.6&77.47&-0.64(7)\\ 
$^{164}Ir$&3.95&-60.8&6.88&-27.0&59.97&-4.06(5)&3.44&-36.2&6.74&-13.5&59.97&-3.97(6)\\ 
$^{165}Ir^*$&3.93&-61.0&6.89&-27.1&62.35&-3.65(5)&3.45&-36.3&6.76&-13.5&62.35&-3.56(5)\\ 
$^{166}Ir$&1.81&-62.1&7.49&-15.1&87.51&-1.18(10)&1.57&-36.8&7.39&-7.6&87.51&-1.13(10)\\ 
$^{166}Ir^*$&3.93&-61.3&6.91&-27.2&80.67&0.07(8)&3.45&-36.5&6.77&-13.6&80.67&0.16(8)\\ 
$^{167}Ir$&0.00&-62.3&7.64&-12.8&91.08&-1.34(8)&0.00&-36.0&7.57&-6.3&91.08&-1.30(8)\\ 
$^{167}Ir^*$&3.94&-61.4&6.92&-27.3&83.82&0.55(8)&3.43&-36.7&6.79&-13.6&83.82&0.64(8)\\ 
$^{171}Au$&0.00&-62.2&7.70&-12.7&69.09&-5.08(15)&0.00&-35.9&7.63&-6.2&69.09&-5.04(15)\\ 
$^{171}Au^*$&3.94&-61.3&7.01&-26.4&64.25&-3.18(4)&3.46&-36.5&6.87&-13.2&64.25&-3.09(5)\\ 
$^{177}Tl$&0.00&-62.5&7.76&-13.2&88.25&-1.44(25)&0.00&-36.1&7.69&-6.4&88.25&-1.39(26)\\ 
$^{177}Tl^*$&3.92&-61.5&7.10&-26.5&57.43&-4.63(6)&3.43&-36.8&6.96&-13.2&57.43&-4.54(6)\\ 
$^{185}Bi$&0.00&-62.7&7.88&-13.1&65.71&-5.52(13)&0.00&-36.3&7.81&-6.3&65.71&-5.47(13)\\ 
 
\end{tabular} 

\end{table}

\section{SUMMARY AND CONCLUSIONS}
\label{section5}

      The half lives for proton-radioactivity have been analyzed with microscopic nuclear potentials obtained by the single folding the DDM3Y effective interaction whose energy dependence parameters have been obtained from nuclear matter calculations. This procedure of obtaining nuclear interaction potentials are based on profound theoretical basis. The results of the present calculations are in good agreement over a wide range of experimental data. It is worthwhile to mention that using the realistic microscopic nuclear interaction potentials, the results obtained for the proton radioactivity lifetimes are noteworthy and are comparable to the best available theoretical calculations. It is therefore observed that the DDM3Y effective interaction provides unified descriptions of cluster radioactivity \cite{r18}, scatterings of $\alpha$ and heavy ions \cite{r11} when used in a double folding model, and nuclear matter \cite{r7,r12} and elastic and inelastic scattering of protons \cite{r8} when used in a single folding model. We find that it also provides reasonably good description of proton radioactivity.

\end{document}